\documentclass[aps,pre,reprint,superscriptaddress,showpacs,floatfix,citeautoscript]{revtex4-1}
\usepackage[latin1]{inputenc}
\usepackage{color}
\usepackage{dcolumn}
\usepackage[dvips]{graphicx}
\usepackage{fixmath}
\usepackage{xspace}
\usepackage[colorlinks=true,linkcolor=black,%
citecolor=black,urlcolor=black,pdfborder={0 0 0}]{hyperref}

\usepackage{amssymb}
\usepackage{amsmath}
\usepackage{amsthm}
\usepackage{bbm}
\usepackage{bm}
\usepackage{multirow}
\usepackage{psfrag}
\usepackage{mathrsfs}
\usepackage{footmisc}

\newcommand{\fett}{\mathbf}
\newcommand{\ve}{\bm}

\newcommand{\te}[1]{\fett{\underline{#1}}}

\newcommand{\etal}{\emph{et al.}\xspace}
\usepackage{subfigure}

\newcolumntype{d}{D{.}{.}{3}}

\begin{document}
\title{Kinetic Activation Relaxation Technique}

\author{{Laurent Karim} \surname{B\'{e}land}}
   \email{laurent.karim.beland@umontreal.ca}
  \affiliation{D\'{e}partement de Physique and Regroupement  Qu\'{e}b\'{e}cois sur les Mat\'{e}riaux de Pointe (RQMP),  Universit\'{e} de Montr\'{e}al, C.P. 6128, Succursale Centre-Ville, Montr\'{e}al, Qu\'{e}bec, Canada H3C~3J7}
\author{{Peter} \surname{Brommer}}
   \email{peter.brommer@umontreal.ca}
  \affiliation{D\'{e}partement de Physique and Regroupement  Qu\'{e}b\'{e}cois sur les Mat\'{e}riaux de Pointe (RQMP),  Universit\'{e} de Montr\'{e}al, C.P. 6128, Succursale Centre-Ville, Montr\'{e}al, Qu\'{e}bec, Canada H3C~3J7}
\author{{Fedwa} \surname{El-Mellouhi}}
\email{fadwa.el\_mellouhi@qatar.tamu.edu}
 \affiliation{Science Program, Texas A\&M at Qatar, Texas A\&M Engineering Building, Education City, Doha, Qatar}
\author{{Jean-Fran\c{c}ois} \surname{Joly}}
   \email{jean-francois.joly.1@umontreal.ca}
   \affiliation{D\'{e}partement de Physique and Regroupement  Qu\'{e}b\'{e}cois sur les Mat\'{e}riaux de Pointe (RQMP),  Universit\'{e} de Montr\'{e}al, C.P. 6128, Succursale Centre-Ville, Montr\'{e}al, Qu\'{e}bec, Canada H3C~3J7}
\author{{Normand} \surname{Mousseau}}
   \email{normand.mousseau@umontreal.ca}
   \affiliation{D\'{e}partement de Physique and Regroupement  Qu\'{e}b\'{e}cois sur les Mat\'{e}riaux de Pointe (RQMP),  Universit\'{e} de Montr\'{e}al, C.P. 6128, Succursale Centre-Ville, Montr\'{e}al, Qu\'{e}bec, Canada H3C~3J7}

\date{\today}

\begin{abstract} 

  We present a detailed description of the kinetic Activation-Relaxation
  Technique (k-ART), an off-lattice, self-learning kinetic Monte Carlo algorithm
  with on-the-fly event search. Combining a topological classification for local
  environments and event generation with ART nouveau, an efficient unbiased
  sampling method for finding transition states, k-ART can be applied to complex
  materials with atoms in off-lattice positions or with elastic deformations
  that cannot be handled with standard KMC approaches. In addition to presenting
  the various elements of the algorithm, we demonstrate the general character of
  k-ART by applying the algorithm to three challenging systems: self-defect
  annihilation in c-Si (crystalline silicon), self-interstitial diffusion in Fe and structural
  relaxation in a-Si (amorphous silicon).
  
\end{abstract}

\maketitle

\section {Introduction}

Solid-phase diffusion in materials science and condensed matter is dominated by
rare atomic diffusion events associated with high-energy barriers as measured
with respect to temperature. These stochastic processes take place on an
extended time scale that makes them very difficult to reproduce using linear
simulation schemes such as molecular dynamics. Low rates, however, allow us to
consider this series of processes as independent Markov chains. In this case, it
is possible to apply the kinetic Monte Carlo (KMC) algorithm proposed by Bortz
\emph{et al.}~\cite{Bortz:1975:10, Fichthorn:1991:1090,
Voter:2002:321,Voter:2007:RES:1}. Based on transition state theory, KMC uses a
catalog of pre-specified diffusion mechanisms to compute at every stage the exit
rate from a local minimum. The clock is then advanced using Poisson's law, a
move is selected with the appropriate rate, and a dynamical trajectory is
constructed. While time steps in KMC are dominated by the lowest energy
barriers, it is possible, under the right conditions, to simulate on experimental
time scales. 

Since it was proposed, KMC has been used extensively in materials science,
condensed matter, and many other fields. Its well-known limitations have
nevertheless prevented the method from being applied widely to complex systems.
In particular, since KMC depends on a predefined event catalog, systems under
study have to be discretized and atomic motion limited to fixed lattice
positions \cite{Voter:2002:321}. In this way, it is possible to evaluate from
the onset all possible moves that will be included in the catalog. For
relatively simple kinetics, such as metal-on-metal growth, these limits are
not major hurdles. They become a problem when trying to take full account of the
lattice associated with long-range elastic effects and, more importantly,
off-lattice and disordered conformations.

Over the years, a number of algorithms have been proposed to lift, at least
partially, these limitations. Most can be classified in one of two categories:
the addition of a continuum approximation for computing the effects of
long-range strain deformations on the energy barriers and on-the-fly evaluation
of energy barriers. The first class of methods adds long-range contributions,
computed from a number of extrapolation schemes, to the energy barriers
extracted from the predefined catalog of
events~\cite{Mason:2004:S2679,Sinno:2007:5}. The second class relaxes the need
for a predefined catalog and, in some cases, moves away from a lattice-based
description. This is the case, for example, of the self-learning KMC approach by
Trushin {\em et al.}, which keeps lattice-based displacement, but introduces an
on-the-fly search for barriers \cite{Trushin:2005:115401}. To remove the
constraints of a lattice-based description, other methods construct a new
catalog at each step to determine the next step, using various climbing methods
such as the dimer~\cite{Henkelman:1999:7010,Hontinfinde:2006:995},
eigenvector-following~\cite{Middleton:2004:8134}, or the autonomous basin
climbing methods~\cite{Fan:2011:125501}. 

The first class of
methods~\cite{Mason:2004:S2679,Sinno:2007:5,Trushin:2005:115401} remains limited
to on-lattice positions in addition to providing often ill-controlled
corrections to energy barriers modified by elastic deformation. The second class, with an
on-the-fly, off-lattice approach, is much more flexible. However, it is
inefficient as a catalog must be rebuilt at each step, making it costly to study
complex systems with a large number of possible diffusion
mechanisms~\cite{Henkelman:1999:7010,Hontinfinde:2006:995,Middleton:2004:8134,Fan:2011:125501}.

In 2008, we introduced the kinetic activation-relaxation technique (k-ART), an
on-the-fly, off-lattice KMC method that lifts these
limitations~\cite{El-Mellouhi:2008:153202}. In this initial work, we showed
that, for a system of vacancies in Si at 500 K, it achieves significant
speed-ups over standard MD, while retaining a complete description of the
relevant physics, including long-range elastic interactions. More
recently, Kara \emph{et al.} proposed a similar self-learning kinetic
Monte Carlo method based, however, on less controlled procedures for finding
barriers and classifying off-lattice
configurations~\cite{Kara:2009:084213}.  

Here, we present in detail the k-ART algorithm, which couples the
activation-relaxation technique (ART nouveau)~\cite{Barkema:1996:4358,
Malek:2000:7723} for generating events and calculating barriers with {\sc
nauty}~\cite{McKay:1981:45} for the topological classification of events. We
also present improvements introduced for handling low-energy barriers and large
systems. Finally, we demonstrate the efficiency and versatility of the method by
applying it to three systems: self-defect annihilation in crystalline Si,
interstitial diffusion in Fe and relaxation in amorphous silicon.

\section{Overview of Kinetic ART}

Following standard KMC, the k-ART method uses an event catalog to compute the
rate of escape from a local minimum and bring forward the simulation clock.
There are three fundamental differences with respect to standard KMC, however.
First, discretization of the local environment is done through topology instead
of geometry, allowing atoms to adopt freely any spatial arrangement instead of
being constrained to predefined lattice positions. Second, the catalog is not
fixed at the simulation onset, but grows as new local environments are visited,
allowing the study of very complex systems. Third, event energy barriers are
fully relaxed at each step to take into account all geometrical rearrangements
due to short- and long-range elastic deformations.

A simulation starts from an initial configuration relaxed into a local energy
minimum. The local topology associated with each atom is first characterized
with {\sc nauty}~\cite{McKay:1981:45}. All atoms sharing a specific topology are
presumed to be associated with the same list of activated mechanisms. This is
the basic assumption of k-ART and, as we will see below, it can be made to hold
on a per-atom basis. This approach results in a considerably reduced amount of
generated and handled data. For a single vacancy in crystalline
silicon (c-Si), for example,
one only needs to consider 20 different topologies to describe all local
environments, irrespective of the system size.

A search for activated pathways is launched for each topology using ART
nouveau~\cite{Barkema:1996:4358,Malek:2000:7723}. ART nouveau was shown to
identify efficiently relevant diffusion mechanisms in systems described with
both empirical and \emph{ab initio} methods~\cite{Song:2000:15680,
Valiquette:2003:125209, El-Mellouhi:2004:205202, Malouin:2007:045211}. This
method has been extensively characterized in Ref.~\cite{Marinica:2011:094119}. A
number of technical improvements are also reported in
Ref.~\cite{Machado-Charry:2011:034102} and new events can now be generated with as
little as 300 force evaluations with either empirical or \emph{ab initio}
potentials. Each event is classified according to the initial minimum, the
saddle configuration and the final state topologies and stored in the catalog.

Once the extensive search for events on all topologies is finished, relevant
events for the current configurations are collected. All low-energy barrier
events are relaxed for specific atoms, to include not only topological but also
geometrical effects. At this point, following Bortz \emph{et
al.}~\cite{Bortz:1975:10}, the elapsed time to the next event is computed as
$\Delta t = -\ln \mu / \sum_i r_i$ where $\mu$ is a random number in the $[0,1[$
interval and $r_i$ is the rate associated with event $i$. The clock is pushed
forward, an event is selected with the proper weight, and the atoms are moved
accordingly, after a geometrical reconstruction.

Once in this new configuration, the process starts again: the topology of all
atoms belonging to the local environment around the new state is constructed; if
an unknown topology is found, a series of ART nouveau searches are launched,
otherwise, we proceed to the next step. After all events are updated, the
low-lying barriers are, once again, relaxed before applying the KMC algorithm.

\section{Algorithmic and implementation details}

\subsection{ART nouveau}

The search for activated mechanisms is performed using ART
nouveau~\cite{Barkema:1996:4358,Malek:2000:7723}, an open-ended climbing method
for finding first-order saddle points surrounding a local minimum. As the most
recent version of the algorithm is described in
Refs.~\cite{Marinica:2011:094119} and \cite{Machado-Charry:2011:034102}, we give here
only a brief overview of the method. Event search with ART nouveau proceeds in
three steps: (1) starting from an energy minimum, the system is deformed locally
in a random direction until the lowest curvature of the Hessian matrix becomes
negative, indicating an instability; (2) the configuration is then pushed along
this direction of negative curvature while the energy is minimized in the
hyperplane orthogonal to this direction until the total force falls below a set
threshold, indicating that a first-order saddle has been reached; (3) the
configuration is pushed over this point and is relaxed into a new minimum. This
set of three configurations---initial minimum, saddle point, and final
minimum---forms an event. 

Since activated processes are local in nature, each event is initiated by
displacing a given atom and its neighbors in a random direction. The exact size
for this displaced region depends on the system studied. In
semiconductors, it involves typically first and second nearest-neighbors. In the
case of Si vacancies, for example, the displacements are applied to the central
atom and all atoms within a 3.0 \AA\ radius from it. The initial convergence
criterion for the saddle point is typically set to 0.5 eV/\AA. This is
associated with the generation of the generic 
event catalog (see below) . Further relaxation
associated with the calculation of specific events uses a 0.1 eV/\AA\ threshold.

To decrease computational cost, ART nouveau never computes the Hessian directly,
but rather uses a mixture of Lancz\'os~\cite{Lanczos:1988:AA} and
DIIS~\cite{Pulay:1980:393,*Shepard:2007:2839} methods for converging to the
saddle point. As discussed in Ref.~\cite{Machado-Charry:2011:034102}, less than 300
force evaluations are generally needed to converge to a first-order saddle
point. Taking into account all processes, including non-converging steps and
relaxation into a new minimum, about 600--800 force evaluations are required, on
average, per successful event search. Relaxation of specific events, which
starts near a reconstructed saddle point, are normally much faster,
necessitating typically 1 to 80 force evaluations. 

\subsection{Topological classification and event generation}
\label{sec:catalog}

Topological classification of the local atomic environment is a crucial step
in k-ART, as it provides a means of discretizing and cataloging local
configurations, while taking into account all possible atomic arrangements and
elastic deformations.

\begin{figure}
        \centering
                \includegraphics[height=5.5cm]{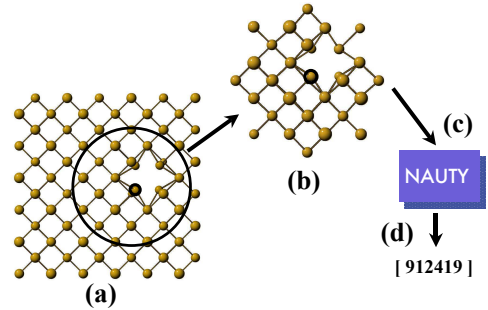}
                \caption{(Color online) Local topology analysis procedure in k-ART. A
                  truncated graph (b) is extracted from the complete
                  lattice (a). This graph is analyzed through {\sc
                    nauty} (c), which returns a unique key and the
                  associated topology (d). }
        \label{fig:topo}
\end{figure}

Atomic topologies for a given local configuration are computed as follows. We define a
local environment consisting of all atoms within a sphere of a predefined radius centered
around each atom, as illustrated in Fig.~\ref{fig:topo}. These are then connected
following a neighboring prescription, such as first neighbor distance cut-off or a Vorono\"{\i}
tessellation, forming a truncated connectivity graph, that is, a set of
bonds connecting vertices, without geometrical information. This graph
is then analyzed and classified 
using the freely available topological software {\sc nauty}, developed by
McKay~\cite{McKay:1981:45}. This software package provides the topology index and all
information necessary for uniquely identifying each environment, including the permutation
key needed to reconstruct a specific geometry from the generic topology and a set of
reference positions. The topology index from {\sc nauty} has the form of an ordered set of
three integers, which we use as input to a hash function to generate a unique topology
label (hash key). 

The geometrical reconstruction from a purely topological graph is made
possible because we know the atomic positions of all atoms surrounding
the local configuration described by this graph. This introduces
sufficient constraints to ensure that most of time, as discussed
below, a given graph corresponds to a unique fully relaxed geometry. 

The neighboring prescription and the size of the truncated region are selected
to ensure that, in most cases, the configuration is uniquely defined
through this network, that is, the connectivity graph must lead to a unique
structure once relaxed with a given interatomic potential. In the case of
crystalline Si, for example, we define the local environment around an atom by a
sphere of radius 5.0 \AA, which includes about 40 atoms; two atoms are linked if
their distance is less than 2.8 \AA.

Once all new topologies are identified, a succession of ART nouveau searches are
launched on each of them. The optimal number of searches done per topology
depends on the nature of the system and can be adjusted to ensure that all
low-energy barrier events, which dominate the dynamics, are found. In our current implementation, it is increased
also with the number of times a given topology is seen, ensuring that the most
common topologies are explored more often.

Every new event found with ART nouveau is analyzed and compared to the
list of already known events for that initial topology. If an event
with the same activation energy is already in memory, a series of
tests are performed to assess whether or not it is the same event. When the generated event is judged to be new,
it is then assigned to the topology centered on the atom that moved
the most during the event, irrespective of the initial activation. The
event label is obtained by combining the topology labels at the
initial, saddle, and final states. We also keep in memory the position
of the cluster of atoms for the initial, saddle, and final
configurations. These are necessary (a) as a reference for the
geometrical reconstruction from the saddle or the final configuration
topological mapping for atoms belonging to the same topological class
and (b) to compare the event in the list with newly generated
events. Once a generic event is added to the database, it is also
added to the binary tree of events and to the histogram (see
Sec.~\ref{sec:histo}).

Finally, based on these data, a first generic rate is associated with the event
by setting:
\begin{equation}
  \label{eq:rate}
r_i = \tau_0
\exp\left(-\Delta E_i/k_BT\right),  
\end{equation}
where $\tau_0$, the attempt frequency, is fixed at the onset and, for
simplification, assumed to be the same for all events ($10^{13}$ s$^{-1}$).
$\Delta E_i$ is the barrier height, that is, the energy difference between saddle
point and initial minimum. $k_B$ and $T$ are, as usual, the Boltzmann
constant and the temperature, respectively. 

While using a fixed attempt frequency is a
  simplification, it has been shown in several systems  that $\tau_0$ varies only
  weakly with the chosen pathway \cite{Yildirim:2007:165421}. The
  value of $10^{13}$ s$^{-1}$ is compatible with pre-exponential
  factors in iron derived from experiment \cite{Stroscio:1994:8522} and
  in simulation \cite{Papanicolaou:2009:1366}. For silicon, this value
  is also compatible with \emph{ab initio} computations of neutral vacancy diffusion
 \cite{El-Mellouhi:2004:205202}.

In certain symmetric configurations, it is possible that distinct events
associated with a given atom have identical topologies of initial state, saddle
point, and final state, and the same barrier height and absolute displacement. To
distinguish between those and fully account for the symmetries, we also include
the direction of atomic motion in the description of events. Checks on
a number of highly symmetric states in Si and Fe show that this
classification recovers all pathways and accurately discriminates
between them.  

Hash keys provide a fast way of storing and retrieving event or topological
information. If the key of a new event or topology is already in use (a so-called hash
collision), the key value is incremented by one until a free key is found. The
arrays for events and topologies account for most of the memory used in k-ART
simulations, and so a balance has to be struck between size and speed. Too large
arrays waste RAM, while too small ones provoke frequent hash collisions and fill
up earlier.

One of the major advantages of this approach is that this constantly updated
catalog of generic events and topologies can be saved to the disk, made
available to others, and reused on the same and similar systems. Multiple
catalogs from different k-ART simulations can also be merged to create a larger
database that can be used to start new simulations on the same system with a
significant speed increase.

\subsection{Adaptive generic event relaxation into specific events}
\label{sec:histo}

Events generated for a given \emph{topology} are known as \emph{generic events}.
It is assumed that all atoms sharing the same topology will have access to these
events with, however, a small adjustment to the energy barrier due to local
variations in position or long ranged elastic interactions. To take these 
changes into account, generic events with low barriers are re-converged for each
realization, resulting in \emph{specific events}, each linked to a particular atom. 

Starting from a common topological generic event, a specific event is generated
for each atom having the same topology by taking advantage of the ordered list
of cluster atoms around the central atom obtained using {\sc nauty} (permutation key).
With this information, it is possible to reconstruct the geometry of the
specific saddle and final minimum conformations by mapping the displacement
vector and transforming geometrically a given region in the system with respect to the generic configuration.

\begin{figure}
        \centering
        \includegraphics[height=7cm]{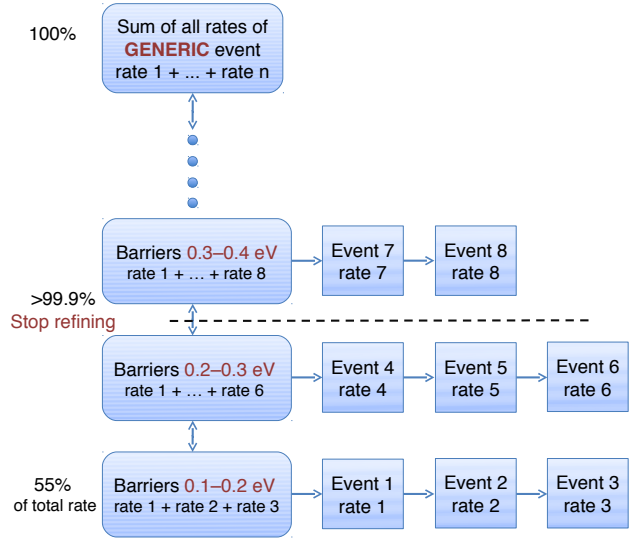}
        \caption{(Color online) Generic events stored in a histogram
          used for an adaptive event relaxation procedure.}
        \label{fig:histo}
\end{figure}

In an earlier version of the algorithm~\cite{El-Mellouhi:2008:153202}, specific
events were identified as those belonging to generic classes with a an
energy barrier of 15 $k_BT$ or less. Here, we adopt rather an adaptive algorithm
based on the kinetics of the system. All generic events are ordered and
stored into slices of 0.1 eV according to their energy barrier in a histogram
(see Fig.~\ref{fig:histo}). For each slice, the total rate is computed
and the cumulative rate up to each energy barrier range is calculated.

The relaxation into specific events starts from the bottom of the histogram and
then proceeds to higher energy barriers until events accounting for a large
fraction of the total rate are relaxed (we use 99.9 \% here). The remaining
generic events are copied to unrefined specific events. In this way, we ensure
that almost all selected moves will be picked from the list of fully relaxed
individual barriers. Geometric and elastic effects on the energies of local
minima and transition states are therefore fully included in the
rates.

\subsection{Crucial aspects}

\subsubsection{Uniqueness in the correspondence between a geometry and
  topology}

One of the main k-ART advantages resides in the topology-based discretization.
To be valid, this classification requires a unique correspondence between a
local geometry and a topology. With the right building rules for the truncated
graph, it turns out that this approximation almost always works for covalently
bonded systems, but also for metals. Since k-ART relies on the delicate
reconstruction of transition states for computing specific events, a failure of
the relation between topology and geometry leads systematically to the
disappearance of this first-order saddle point and can be easily detected. In
other words, event reconstruction automatically fails when a given topology
corresponds to more than one geometry.

When an ill-defined topology corresponding to more than one geometry is
encountered for the first time, k-ART automatically adjusts the atom-atom
connectivity cutoff used within the cluster of atoms forming this graph, until
the geometries initially associated with the same topology are now placed into
two different classes. The unique correspondence between local geometry and
topology is now re-established.

For computational ease, a marker identifies already encountered difficult
topologies so that k-ART can recognize them on the fly and test them to ensure
that the correct event catalog is used. To do so, k-ART maps one event from
each of the associated sub-topologies, until a successful map is achieved. If no
event can be mapped using the current sub-topologies, a new label is again
issued. Each difficult topology can have as many sub-topologies associated with
it as necessary. In practice, topology collisions are extremely rare,
if reasonable cluster and neighbor cutoffs are used. If there is a
collision, it can  usually be resolved with only one sub-topology. 

\subsubsection{Handling low-energy barriers}
\label{sec:lowbarr}

In KMC, dynamics is dominated by a system's lowest barrier energy. When the
energy landscape consists of basins with numerous states connected by
very low-energy barriers compared to those needed to leave these
basins, the algorithm 
becomes trapped into computing non-diffusive events, decreasing significantly
its efficiency in two ways. First, it limits the attainable simulated time, as
the low-energy internal barriers produce a high total rate sum and thus a short
average KMC time increment. Second, computational resources are bound to explore
the states within a basin without yielding much information, as
effective diffusion takes place typically outside the energy basins. 

We had previously implemented a TABU-like approach~\cite{Glover:1997:TS:549765},
that bans transitions rather than states \cite{Chubynsky:2006:4424}. This
algorithm is simple to implement and provides a thermodynamical solution when
there is a clear energy separation; it fails, however, when few pathways are
available or the energy spectrum is continuous. To account for these situations,
we developed the basin-autoconstructing mean rate method (bac-MRM), a
basin-based acceleration scheme inspired by the mean rate method (MRM) of
Puchala \etal \cite{Puchala:2010:134104}. A description of MRM can be found in
the Appendix. In summary, MRM separates the trajectory into transient states and
absorbing states, and accelerates the simulation by averaging over all possible
jumps between transient states, yielding the correct probability to exit a basin
to a certain absorbing state.

In kinetic ART, the relevant entities are not states, but \emph{events},
characterized by an energy barrier between the initial and final states. In an
event with energies $E_i$, $E_s$, and $E_f$ at the initial state, the saddle
point, and the final state, respectively, we define the forward energy barrier as
$b_f=E_s-E_i$, and the inverse barrier by $b_i=E_s-E_f$. In both cases, rates
going forward or backward are determined by Eq.~\eqref{eq:rate}. Basins are then
identified on the fly by the barrier heights separating the basin states: Both
the forward and the inverse barrier must be smaller than a user-defined
threshold.

Starting in a local minimum, low-energy barrier events are marked as potential
basin events, and the atomic displacement associated with these events is
stored. If such an event is picked for execution, it is added to the current
network of basin events (i.e., for the first basin event, this is at that time
the only event), removed from the tree of available events, and executed as
normal. The system is then
in state two, and the k-ART event finding algorithm is started from
this state. All events from previous basin states are kept in the tree
and could be picked as KMC move.

After state two has been searched for possible events, and before the
next event is picked, the MRM is applied to the basin
consisting of two states: The rates leaving the basin are modified
following Eq.~\eqref{eq:accrate}, and the total rate is adjusted
accordingly. The next step is then selected. It can either lead to a new
basin state, to be added according to the procedure described
above, or out of the basin, in which case a standard k-ART move is
applied. If an event is found to lead to an already explored basin
state, it is rejected, removed from the tree, and added to the basin,
adjusting the rates as needed.

Bac-MRM explores basins on the fly, and only as far as necessary.
Simultaneously, no state is intentionally visited twice. While the internal
dynamics within a basin is lost, the basin mean rate method in our
implementation yields the correct distribution of exit states depending on the
basin internal rates and the point of entry into the basin.

The computational overhead of bac-MRM is small and the CPU
resources needed for all basin related operations are negligible
compared to the time required to explore a single topology.

\subsubsection{Optimizing k-ART for large scale systems}
\label{sec:largesc}

Because activated mechanisms are local in nature, it is relatively
straightforward to optimize k-ART to handle systems with several tens of
thousands of atoms. Indeed, diffusive motion typically involves regions composed
of a few tens to a few hundreds of atoms, and the forces induced by this
displacement typically propagate up to a few nanometers. Therefore, by coupling
standard cell lists \cite{Allen:1988:CSL:53563} and the Verlet algorithm
\cite{Verlet:1967:159} for constructing neighbor lists with a local force
calculation, the computational effort of generating an event becomes almost
system-size independent.

Local forces are first computed on all atoms involved in the event plus their
first and second neighbors. As the system evolves, atoms on which the force
exceeds a set threshold (of 0.01 eV/\AA, in the case of c-Si) become
labeled as \emph{involved} and their first and second neighbors are added to the
list. This process ensures that forces are computed only on the relevant atoms.
The generation of new events and the relaxation of specific events therefore is
entirely local, with only a global minimization performed after each KMC move to
take into account all elastic effects.

\begin{figure}
\begin{center}
\includegraphics[width=8.6cm]{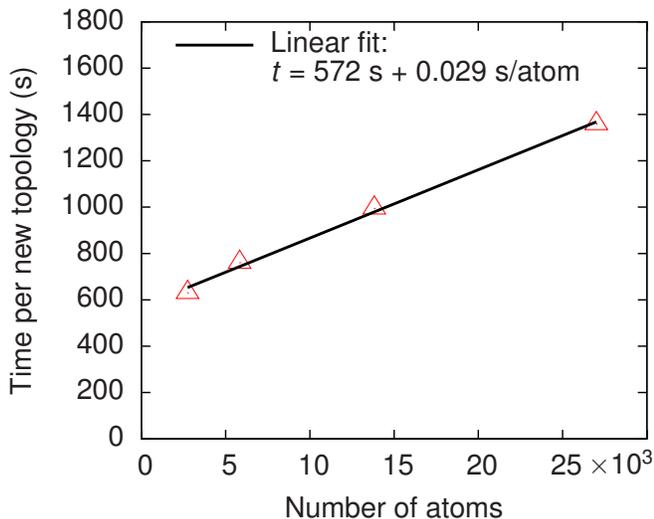}

\caption{(Color online) The average computational time required to explore a new
topology as a function of system size. We show results with 2744 atoms, 5832
atoms, 13\,824, and 27\,000 atoms with 1, 2, 4, and 8 vacancy and interstitial
defects, respectively.}

\label{size_scale_graph}
\end{center}
\end{figure}

We simulated several systems composed of 2744 to 27000 c-Si atoms using
an empirical Stillinger-Weber potential, periodic boundary conditions, and
roughly the same density of vacancies and interstitials in equal numbers. The
scaling is given as a function of system size in Fig.~\ref{size_scale_graph} for
exploring a new topology on a single processor. Because of rare global
calculations during the analysis of events, the algorithm has a weak linear
dependance of 0.03 s per atom. However, the sub-linear terms, responsible for
about 600 s, are dominant for system size of several tens of thousands of atoms.
These sub-linear terms are associated to the time required to attempt 15 saddle
point searches on one topology and to analyze the results as computed
on a single core of a 2.66GHz Intel Xeon X5550 CPU (all CPU times
reported in this paper are computed on the same CPU).

\section{Application of k-ART}
\label{sec:application-k-art}

We now apply k-ART to three different systems, in order to demonstrate the
flexibility of the method: (a) Vacancies and interstitials in c-Si. This
work expands on the vacancy diffusion study presented in
Ref.~\cite{El-Mellouhi:2008:153202}. (b) Interstitials in Fe. Diffusion
mechanisms for self-interstitial in iron are surprisingly complex; this system
represents a good test of k-ART's ability to sample such landscape. (c)
Relaxation of amorphous silicon (a-Si). By construction, KMC methods have been mostly limited
to lattice-based problems; here, we show that the topological approach of k-ART
is sufficiently flexible to handle disordered materials.

\subsection{Vacancies and interstitials in Si}

\begin{figure}
\begin{center}
\includegraphics[width=\columnwidth]{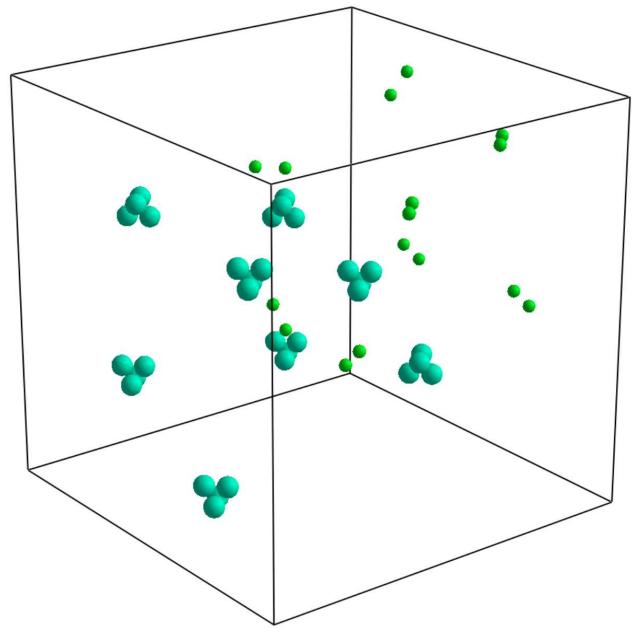}
\caption{(Color online) The initial state of our 8000-atoms c-Si box containing eight
  vacancies and eight interstitials. We only show over-coordinated (small spheres)
  and under-coordinated atoms (large spheres).}
\label{picture_IV}
\end{center}
\end{figure}

We studied the annealing of eight pairs of vacancies and interstitials in an
8000 atoms \emph{c}-Si box at 500 K using the Stillinger-Weber potential
\cite{Stillinger:1985:5262} and periodic boundary conditions. A snapshot of the
initial state of the simulation is shown in Fig.~\ref{picture_IV}. With this
interatomic potential, the diffusion activation barrier for an individual
vacancy is about 0.43~eV \cite{Maroudas:1993:15562}. Interstitials can be
stabilized in three states, two of which are almost degenerate in energy at
about 0.75~eV above the first one. Diffusion for the single self-interstitial is
dominated by a barrier at 0.94~eV \cite{Sinno:1996:3028}.

\begin{figure}
\begin{center}
\includegraphics[height=12cm]{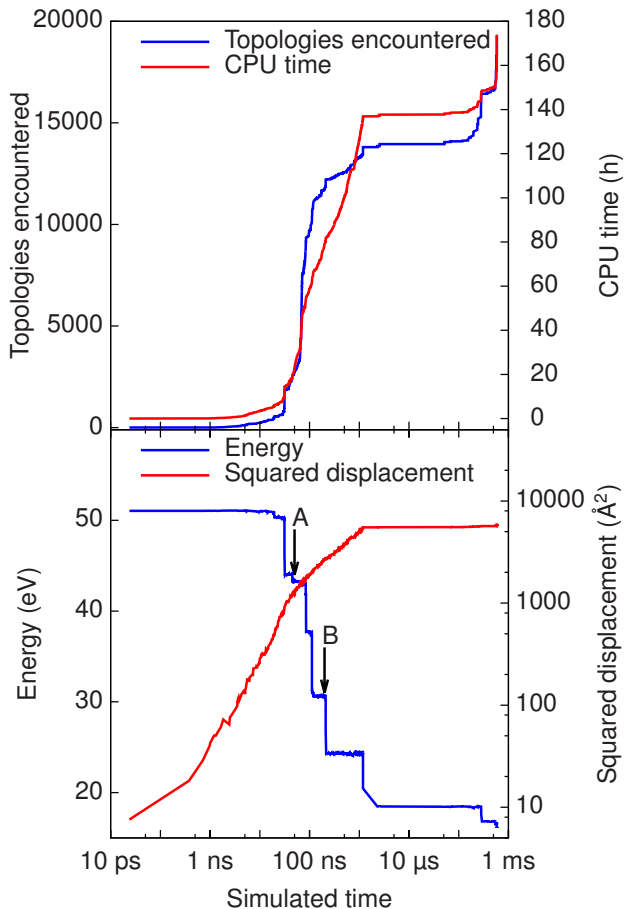}

\caption{(Color online) Simulation of eight IV pairs in an 800-atom box at 500 K. Top: The
number of encountered topologies and the computation time as a function of the
simulation time. Bottom: The evolution of the total energy, measured from the
perfect crystal, and of the the squared total displacement as a function of the
simulation time. The zero on the energy scale corresponds to a box with no
defect. Arrows indicate important interstitial-vacancy annihilation states shown
as snapshots in Figs.~\ref{recomb_IV} and \ref{ocsill_recomb_IV}.}
\label{energy_IV}
\end{center}
\end{figure}

\begin{figure}[]

\subfigure[$\Delta t=0$ ps]{
\includegraphics[width=.4\columnwidth]{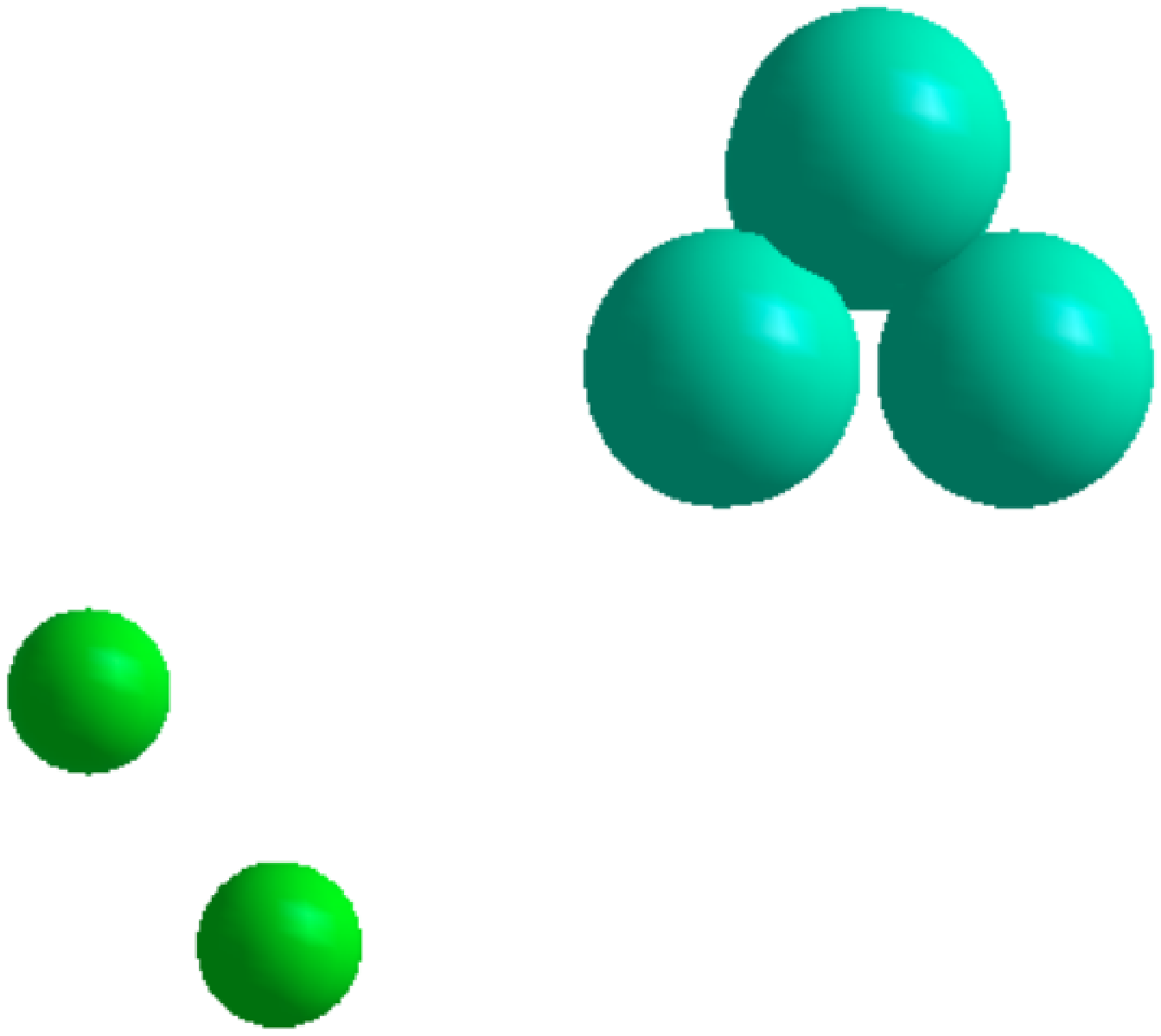}
}
\subfigure[$\Delta t=129$ ps]{
\includegraphics[width=.4\columnwidth]{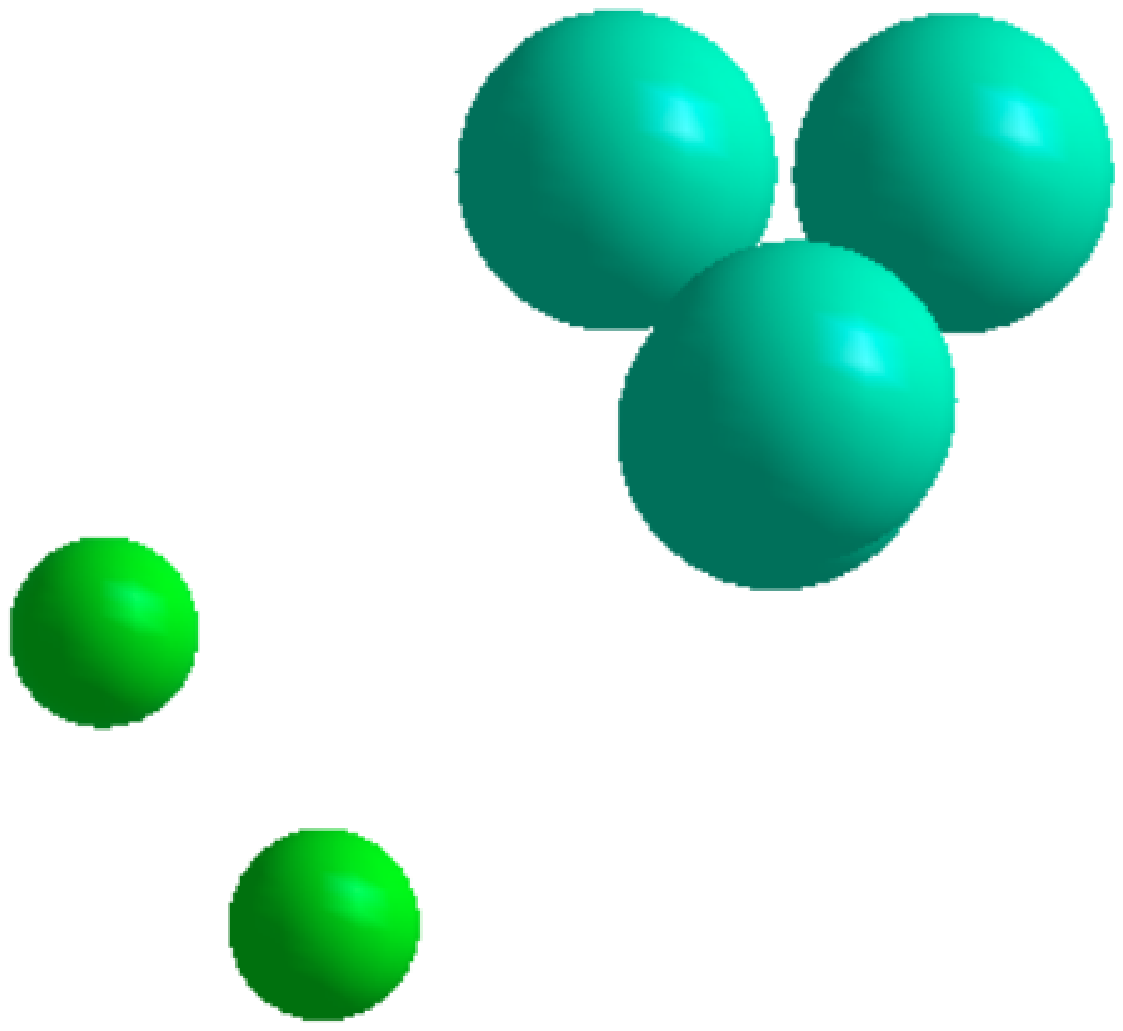}
}
\subfigure[$\Delta t=351$ ps]{
\includegraphics[width=.4\columnwidth]{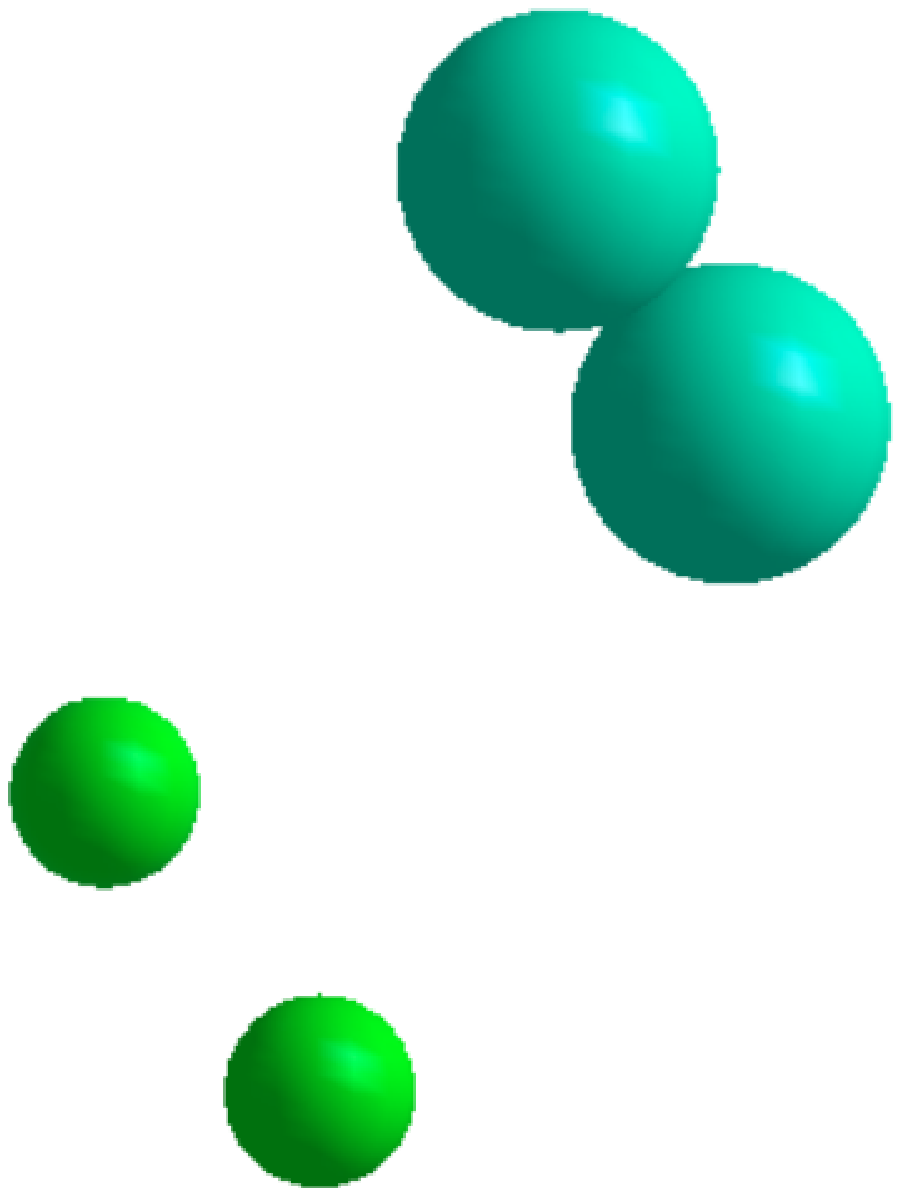}
}
\subfigure[$\Delta t=652$ ps]{
\includegraphics[width=.4\columnwidth]{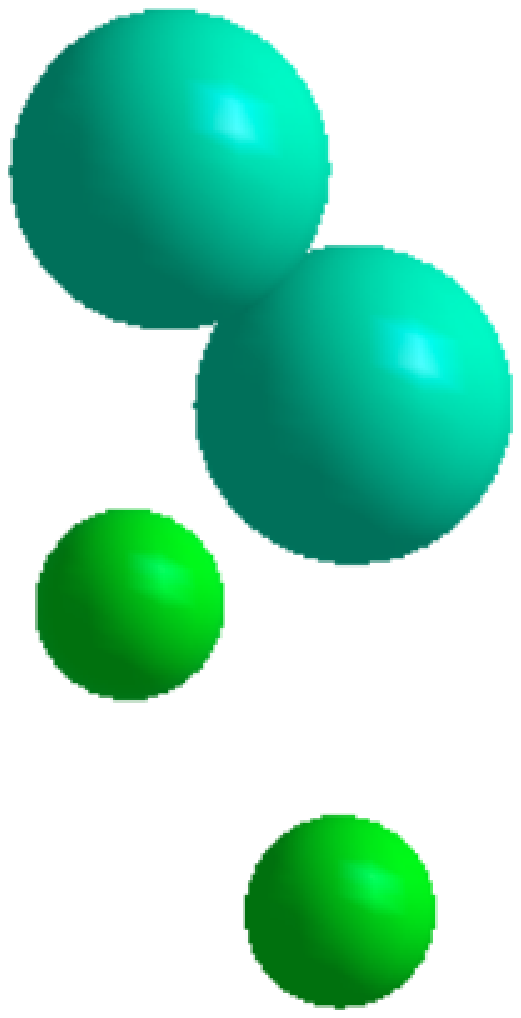}
}
\subfigure[$\Delta t=658$ ps]{
\includegraphics[width=.4\columnwidth]{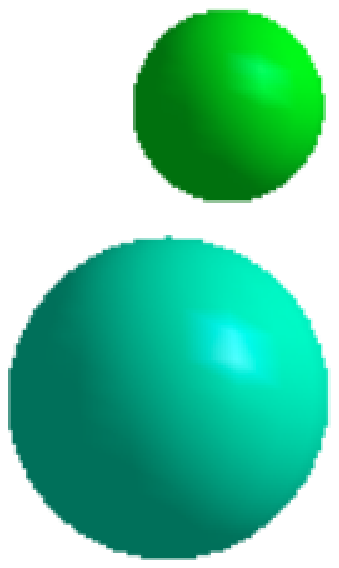}
}
\subfigure[$\Delta t=659$ ps]{
\includegraphics[width=.4\columnwidth]{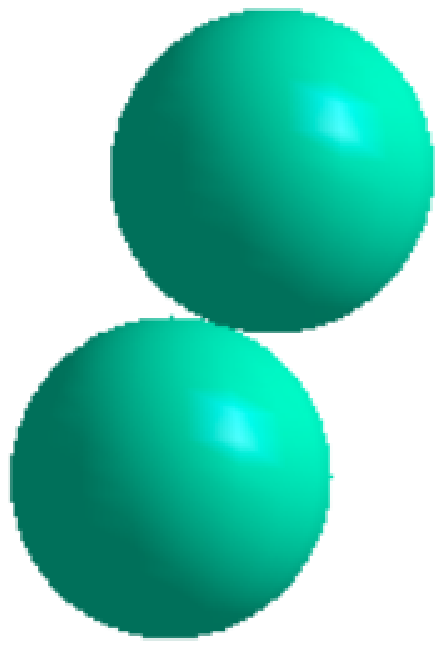}
}
\subfigure[$\Delta t=667$ ps]{
\includegraphics[width=.4\columnwidth]{06g_recomb_IV06.eps}
}

\caption{(Color online) A typical interstitial-vacancy recombination.  Time are measured from the initial snapshop, taken at 0.2 $\mu$s and indicated by arrow B in Fig.~\ref{energy_IV}. Over-coordinated atoms are
shown as small spheres and under-coordinated atoms are shown as large spheres. The empty figure
in (g) indicates that there are no more topological defects in the local
environment.}
\label{recomb_IV}
\end{figure}

Figure~\ref{energy_IV} (bottom) shows the evolution of the total energy,
measured with respect to the crystalline state, and the squared displacement as
a function of simulated time for the system above, representing a total of 2000
k-ART steps. Vacancies dominate the diffusion with significantly lower energy
barriers. Each of the five large drops in energy corresponds to the annihilation
of an interstitial-vacancy (IV) pair, while from roughly 1 $\mu$s on, bound
defects reorganize themselves without any recombination. We show in
Fig.~\ref{recomb_IV} an example of a typical isolated recombination. We observe
several metastable states for the bound pair, before recombination, in general
agreement with the findings of Tang \emph{et al.}~\cite{Tang:1997:14279} and
Marqu\'es \emph{et al.}~\cite{Marques:2001:045214}.

\begin{figure}[]
\subfigure[$\Delta t=0$ ps]{
\includegraphics[width=.4\columnwidth]{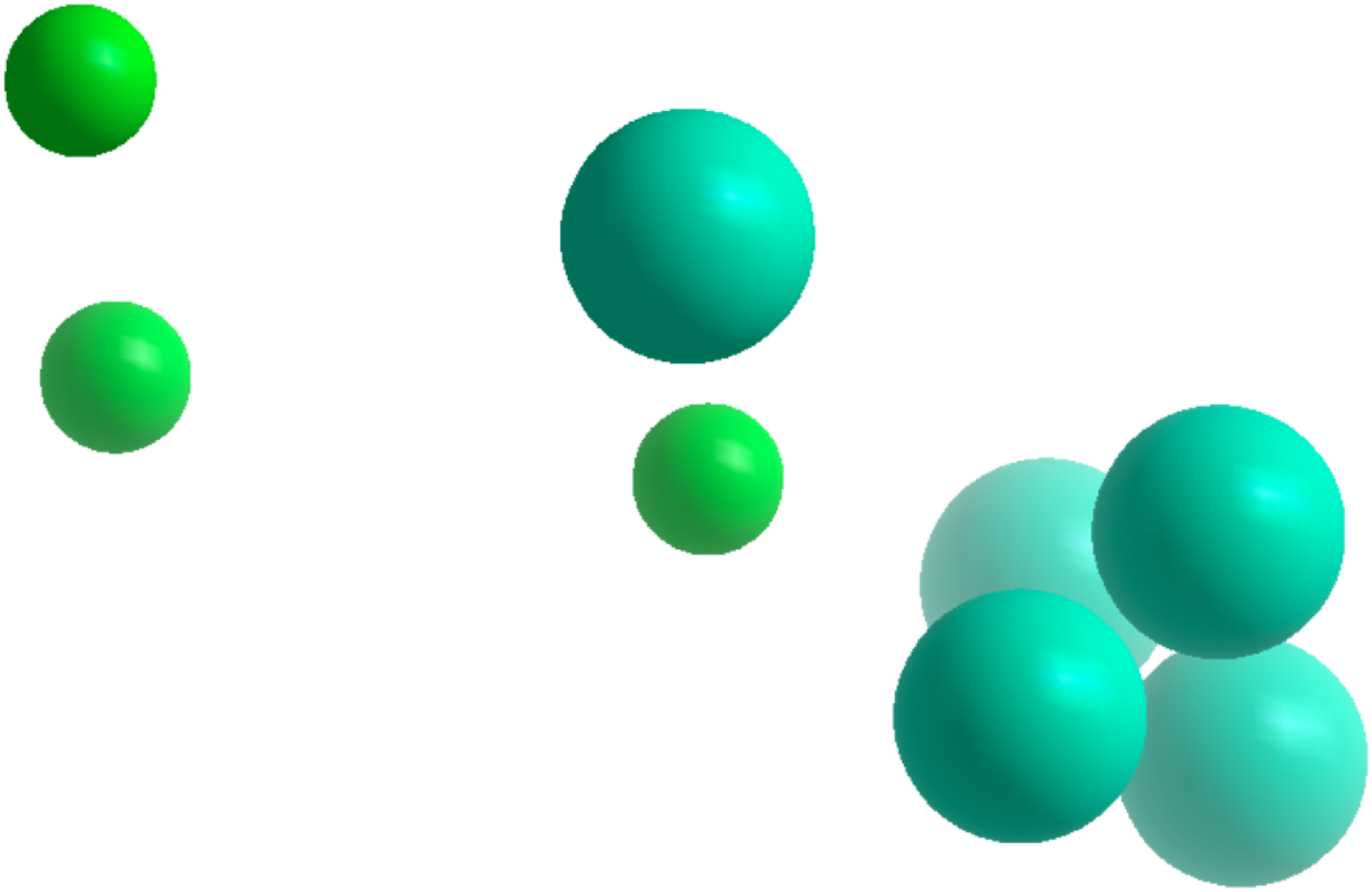}
}
\hspace{10mm}
\subfigure[$\Delta t=150$ ps]{
\includegraphics[width=.4\columnwidth]{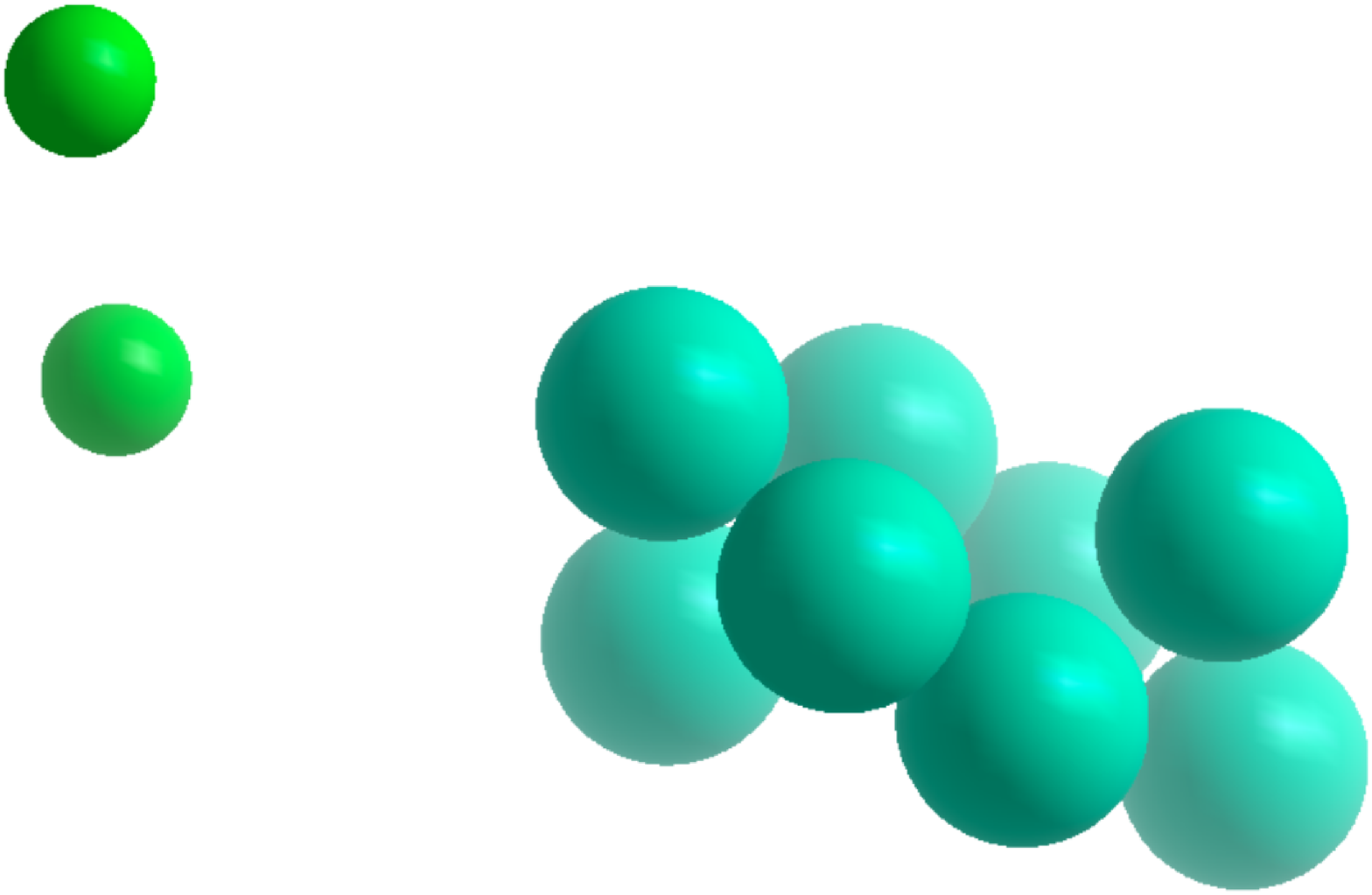}
}
\hspace{10mm}
\subfigure[$\Delta t=152$ ps]{
\includegraphics[width=.4\columnwidth]{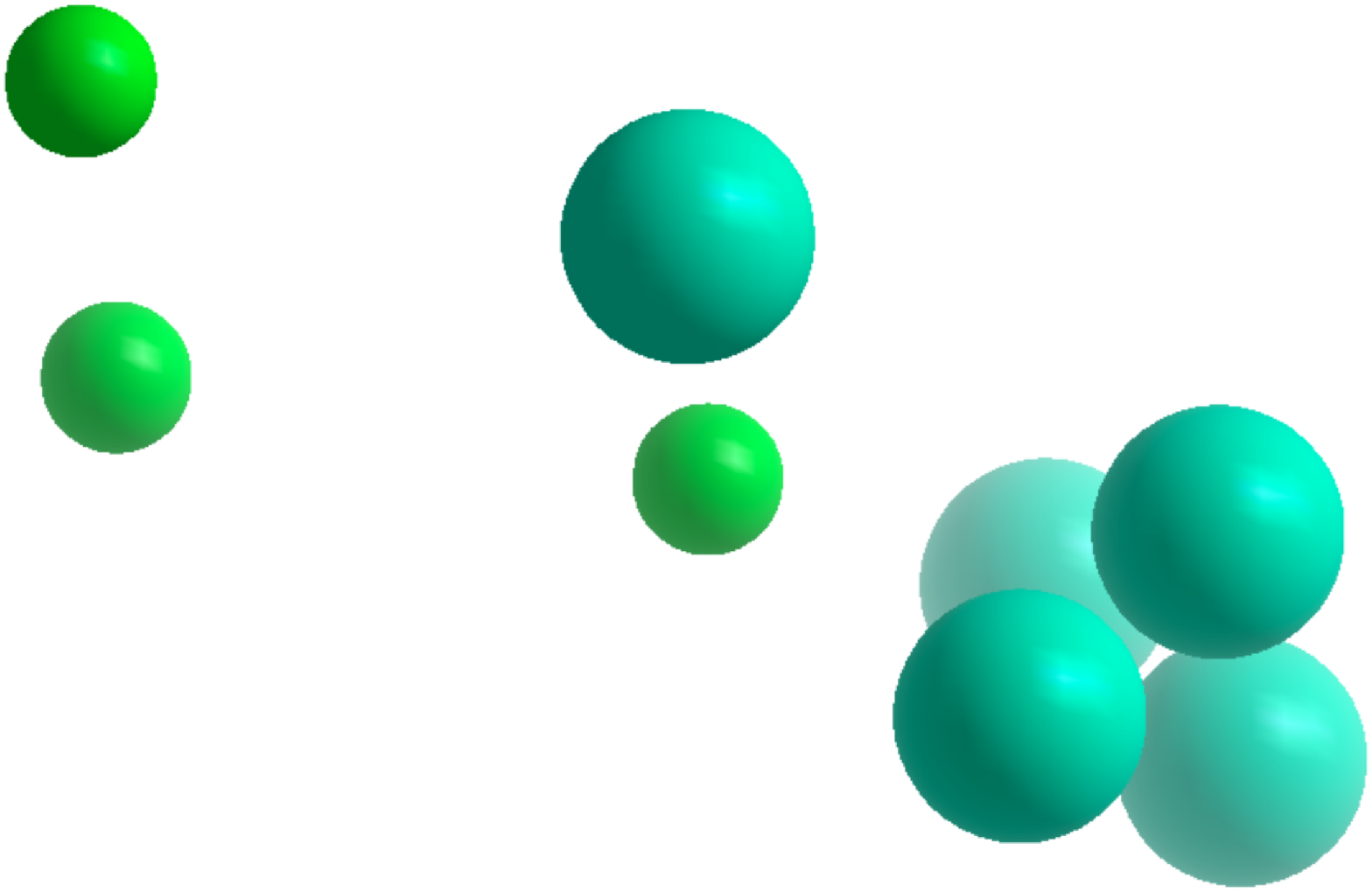}
}
\hspace{10mm}
\subfigure[$\Delta t=322$ ps]{
\includegraphics[width=.4\columnwidth]{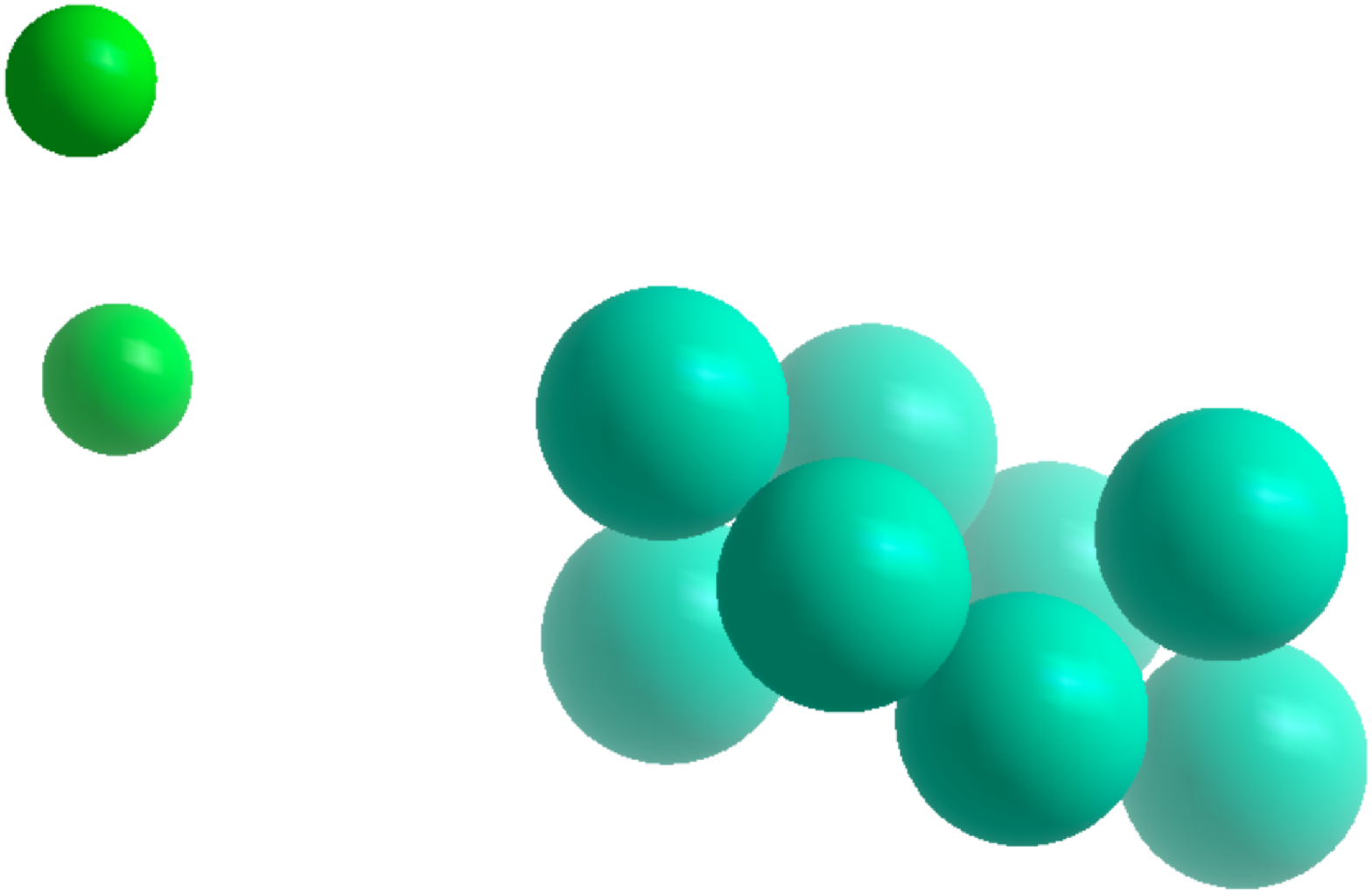}
}
\caption{(Color online) Oscillation of an interstitial-divacancy complex during
an interstitial-vacancy recombination in the presence of nearby
vacancies. Time is measured from the initial snapshot, taken at 50 ns
and indicated by arrow A 
in Fig.~\ref{energy_IV}. Over-coordinated atoms are shown as small spheres and
under-coordinated atoms are shown as large spheres.}
\label{ocsill_recomb_IV}
\end{figure}

Energy fluctuations in Fig.~\ref{energy_IV} are also associated with the
formation of self-defect aggregates, such as a bi-interstitial and a
vacancy complex, as well as elastic deformations. These can cause important
differences between IV-pair recombinations. This is the case of the second IV
recombination (at $t= 0.08\;\mu$s), which differs markedly from the four others.
Indeed, the presence of nearby vacancies (at a distance of about 1 nm from the
main interstitial and vacancy) significantly modifies the relaxation mechanism.
They introduce an intermediate oscillatory state which flickers for a duration
of nearly 45 ns (see Fig.~\ref{ocsill_recomb_IV}) before IV recombination.
This metastable state underlines the importance of correctly handling long-range
elastic and topological effects for the defect kinetics in semiconductors.

Even small changes in the environment, up to 20 \AA\ away from the defect, can
change the kinetics of defect diffusion and recombination. For example, IV-pair
recombination, for a nearly, but not completely isolated pair can follow at
least two pathways, with barriers differing significantly: 0.45, 0.39, 0.19, and
0.20 eV, in the first case, which is in good agreement with Gilmer \emph{et
al.}~\cite{Gilmer:1995:247}, and 0.36, 0.19, 0.22, and 0.51 eV, in the second.

The advantage of k-ART's approach to catalog building can be seen in
Fig.~\ref{energy_IV} (top). Since only previously seen topologies are included, the
catalog grows as new regions of the configurational space are visited, avoiding
the need to construct all possible conformations from the onset, a task that
would rapidly be impossible, even for a system with 16 defects. We see that as
the defects first diffuse in isolation for the first 20 ns, very few new
topologies are encountered. It is only as the recombination processes take
place, between $t=20$~ns and 1~$\mu$s, that many new environments are visited,
increasing rapidly the number of sampled conformations from a few hundreds to
almost 20\,000 topologies. While this number is large, all events are stored in a
catalog and serve as such in any new simulation, reducing considerably the CPU
cost over time.

\subsection{Self-interstitial cluster diffusion  in iron}
\label{sec:intiron}

Iron is widely used in nuclear power plants, which makes the modeling of
irradiation-induced defects interesting, in particular single and clustered
self-interstitial atoms (SIA)~\cite{Fu:2005:68}, from the microscopic to large
scale, an important topic in computational materials science
\cite{Massoud:2010:2,Malerba:2010:7}.

Simulating SIAs on an atomistic level is challenging for both molecular dynamics
(MD) and standard KMC methods. On the one hand, the activation energies of
defect migrations are rather high, so that MD simulations must be performed at
comparatively high temperatures (up to 1200 K, cf.\
Refs.~\cite{Terentyev:2007:104108,Anento:2010:025008}) to explore the energy
landscape within the accessible simulation times. This makes it difficult to
identify structures and mechanisms important at the significantly lower
operating temperatures. On the other hand, due to the wealth of arrangements of
interstitial atoms and other defects, it is a formidable task to build a catalog
of possible transitions \emph{a priori} (to use in a KMC simulation), without
accidentally neglecting important migration paths. Often, a very reduced catalog
of transition pathways is used \cite{Becquart:2010:9}. All this is complicated
by the fact that in iron clusters of interstitial atoms can glide in one
dimension with very small migration energies (tens of milli-electron volts)
\cite{Terentyev:2007:104108}.

K-ART is an ideal tool to explore the migration pathways of SIAs without any
assumptions needed to assemble a catalog of events and the associated barriers
\emph{a priori}. Events are searched for on the fly for each topology in the
system, with corrections applied at each KMC step to account for elastic
distortions by surrounding defects. As outlined in Sec.~\ref{sec:lowbarr}, low
barriers can also be treated efficiently with the basin-autoconstructing mean
rate method.

For our iron simulations, we used the Ackland-Mendelev potentials
\cite{Ackland:2004:S2629}, an improved version of the potential developed by
Mendelev \etal\ \cite{Mendelev:2003:3977}. This potential describes defects
accurately~\cite{Malerba:2010:19} and was used in
MD~\cite{Terentyev:2007:104108,Anento:2010:025008} and ART
nouveau~\cite{Marinica:2011:094119} simulations of iron SIA systems. The EAM
energy and force calculation routine was adapted from \textsc{IMD}
\cite{Stadler:1997:1131} and can use any tabulated potential in \textsc{IMD}
form.

To test the implementation, a single self-interstitial atom was embedded in a
1024-atom supercell. In the ground state, the interstitial forms  a dumbbell in
[110] direction in agreement with earlier static simulations
\cite{Fu:2004:175503} and ART studies \cite{Marinica:2011:094119}. The
transition with the lowest energy barrier follows the nearest neighbor (NN)
translation-rotation mechanism proposed by Johnson \cite{Johnson:1964:A1329}
with an activation barrier of 0.3~eV. Higher barrier events include a
transformation to the [11$\xi$] dumbbell, followed by an on-site rotation, then
pure translations to first and second NN sites. Other higher-energy events were
found, but were not picked to be executed during our simulations.

\begin{figure}
\begin{center}
\includegraphics{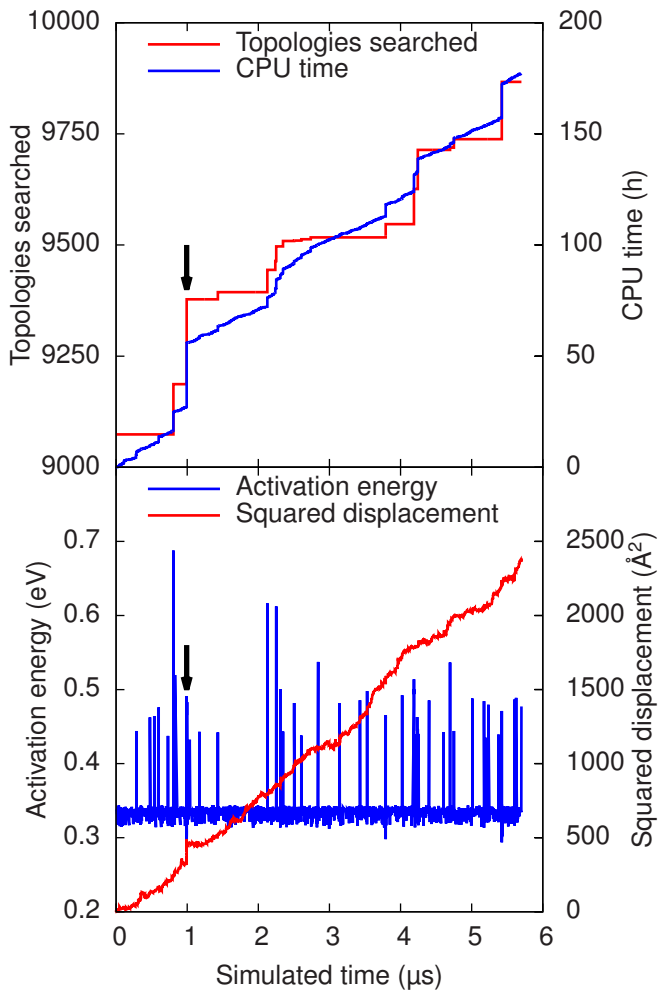}
\caption{(Color online) Simulation of an iron di-interstital cluster in a 1024-atom
  box at 300 K. Top: The
number of encountered topologies and the computation time as a function of the
simulation time. Bottom: The evolution of the total energy, measured from the
ground state, and of the the squared total displacement as a function of the
simulation time. The system propagates mainly
  by the two-step Johnson process with an activation energy around
  0.33 eV. The simulation was performed
  using a pre-built catalog constructed from an earlier simulation,
  initially containing information about 9074 topologies. The
  simulation stalls when many new topologies must be explored. The
  arrow marks the basin shown in detail in
  Fig.~\ref{fig:2iu_basin_detail}. In that basin, the activation
  energy is under 0.3 eV. Two more basins are encountered at 3.8
  and 5.5 $\mu$s.}
\label{fig:2iu_basinplot}
\end{center}
\end{figure}

In simulations with two clustered interstitials, k-ART recovers the mechanism
for interstitial migration suggested by Johnson \cite{Johnson:1964:A1329}: Both
interstitials each perform a nearest-neighbor translation-rotation jump. This
can happen in a single step, or in two sequential moves. The states and barriers
found in our simulation agree with the results from Marinica \etal\
\cite{Marinica:2011:094119}. A sample trajectory of the di-interstitial system
over 2000 KMC steps is shown in Fig.~\ref{fig:2iu_basinplot}.

\begin{figure}
\begin{center}
\includegraphics{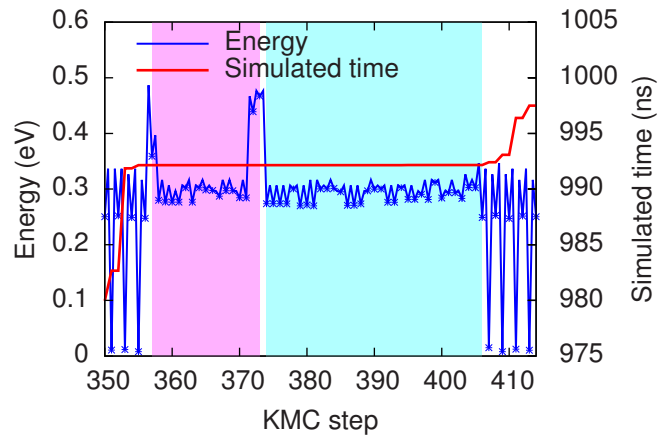}
\caption{(Color online) Detail of Fig.~\ref{fig:2iu_basinplot} (KMC steps 350--413,
  0.99 $\mu$s): After a series of two-step Johnson jumps, the system
  crosses into an excited state at KMC step 235. Then a number of
  states forming a basin are traversed. Shaded background indicates a
  basin (oscillatory) motion. After an exit event, the system resumes its basin
  trajectory, until another exit event leads it back to the ground
  state, from where it resumes its two-step motions. The energy
  trajectory passes through minima and saddle points alternately
  (minima marked by crosses). During the basin motion, the system
  clock is hardly moving.}
\label{fig:2iu_basin_detail}
\end{center}
\end{figure}

For a single self-interstitial, at the temperature of interest, barriers are
well above $k_BT$, and there are no flickers. For SIA clusters, however, there
are sequences of states separated by low barriers. In the di-interstitial case,
Marinica \etal\ \cite{Marinica:2011:094119} find a basin of 0.25 eV above the
ground state, with barriers below 0.1 eV separating the states. We reproduce
that basin as demonstrated in a detail of the trajectory in Fig.\
\ref{fig:2iu_basin_detail}. While the system explores the basin, low-energy
barriers keep the system clock almost at a standstill.

\begin{figure}
\begin{center}
\includegraphics{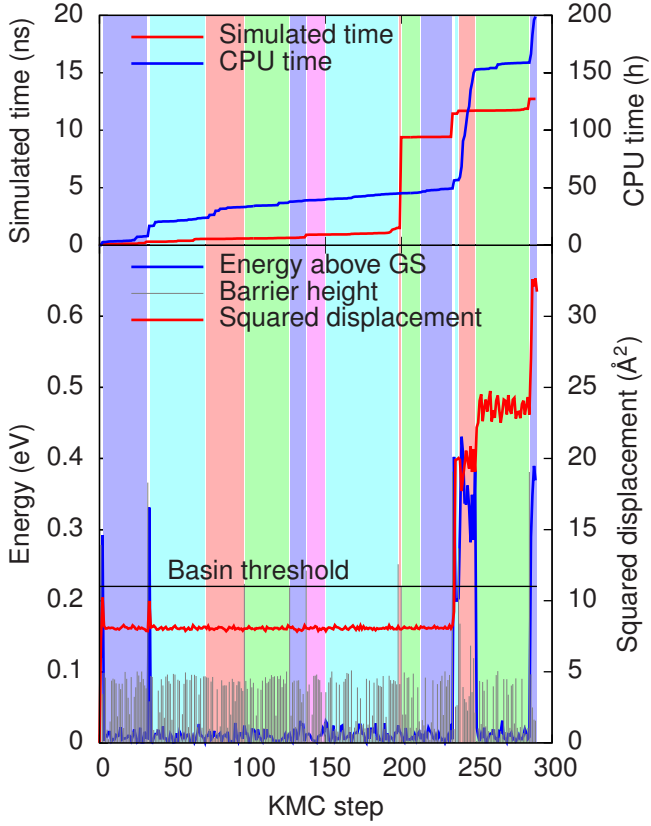}
\caption{(Color online) Trajectory of a tetra-interstitial cluster at 300 K in a
  1024-atom cubic box over 290 KMC steps. Top: The simulation and the computation time as a function of KMC steps. Bottom: The evolution of the total energy, measured from the
perfect crystal, and of the the squared total displacement as a function of KMC steps. The system makes several attempts to leave the
  ground state basin, but falls back until, at KMC step 234, it
  succeeds. It then moves through a sequence of excited states, before dropping
back to a different ground state basin, with the whole cluster
diffusing to the nearest neighbor site. Different background colors represent
different basins (white: outside of basin). As the barriers (shown as
impulses in the lower plot) are
comparatively low, the clock advancement is rather small. Only if
a barrier exceeding the basin threshold is
picked, the KMC time step is noticeable. A significant share of the CPU
time is spent exploring the sequence of excited states between steps
234 and 250.}
\label{fig:4iu_basinplot}
\end{center}
\end{figure}

In a system with a 4-SIA cluster, a basin is found around the ground state:
Small reorientations of the four dumbbells lead to about 20 unique
configurations separated by extremely low barriers ($<$0.1 eV). Since the
dynamics is dominated by low-barrier events, the system manages to exit the
basin only when all those states have been explored. A sample trajectory over
290 KMC steps is displayed in Fig.\ \ref{fig:4iu_basinplot}.

The wealth of structures and transition pathways found in iron systems
with SIA clusters is virtually impossible to include in a catalog
assembled \emph{a priori} for standard KMC simulations. In contrast, the
self-learning k-ART program will over the course of a simulation build
a database of these configurations and events, thus saving time once
the system revisits previous states. The presence of basins
(i.e.~groups of states separated by low barriers) in these
systems dictates the use of an acceleration
scheme. With bac-MRM, even such a rich system becomes a tractable
problem in KMC simulation.

\subsection{Relaxation dynamics in amorphous silicon}

As kinetic ART is an off-lattice method with with self-learning catalog building
capabilities, it can also be used to study relaxation of disordered materials on
long time scales. There have been previous applications of similar techniques to
these systems, but these have suffered from limited sampling of
events~\cite{Mousseau:2001:775,Middleton:2004:8134}.

As a test case, we looked at amorphous silicon. Like crystalline silicon,
this allotrope of silicon is fourfold coordinated with randomly oriented
tetrahedra causing the loss of medium and long range order in the system. This
model system has been extensively studied with
ART~\cite{Barkema:1996:4358,Mousseau:2000:1898} and ART
nouveau~\cite{Valiquette:2003:125209,Kallel:2010:045503} and constitutes
therefore a well-controlled model. Moreover, many fundamental questions remain
regarding its dynamical properties. For example, in spite of considerable
experimental efforts, the exact nature of defects responsible for structural
relaxation is still a matter of
debate~\cite{Roorda:1991:3702,Coffa:1991:2916,Coffa:1992:8355,
Mercure:2005:134205}. As for many other disordered systems, only methods able to
reach experimental timescales will be able to offer a satisfactory answer to
these questions.

\begin{figure}
\begin{center}
\includegraphics{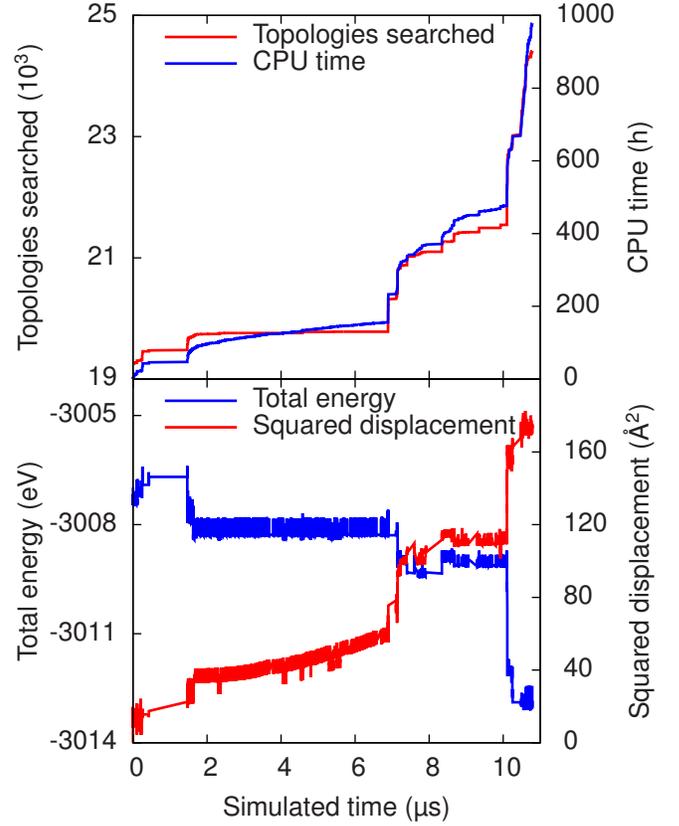}
\caption{(Color online) Simulation of a-Si in a 1000-atom box at 300 K. 
 Top: The
number of encountered topologies and the computation time as a function of the
simulation time. Bottom: The evolution of the total energy and of the the squared total displacement as a function of the
simulation time.
 The simulation
  was started with a catalog from an earlier simulation. The system
  flickers 
between two neighboring states until it finds a way to relax further. This leads
to a sequence of configurations never seen before and the CPU time needed per
step increases with the number of new topologies to explore.}
\label{asi_plot}
\end{center}
\end{figure}

We start here with a well-relaxed 1000-atom a-Si configuration with
periodic boundary conditions generated with the modified Wooten-Winer-Weaire
procedure \cite{Wooten:1985:1392, Barkema:2000:4985} and a reparametrized
version of the popular Stillinger-Weber potential by Vink \emph{et
al.}~\cite{Vink:2001:248} adjusted to describe appropriately this allotrope. All
atoms in the generated a-Si sample are perfectly coordinated with a clean
electronic gap \cite{Durandurdu:2000:15307} and a good agreement with the
experimental radial distribution function~\cite{Barkema:2000:4985}.

For a disordered system, the advantage of recycling events based on the local
atomic topologies takes a lot of time before becoming noticeable. For a
well-relaxed 1000-atom model of a-Si, for example, no two atoms share the
same topology and even after many thousands of events, topologies encountered
more than once are rare. A meaningful catalog requires therefore the combination
of many independent KMC trajectories started from various initial
configurations.

At first, since each atom has its own topology, the number of initial events to
be generated is very large. Successive steps tend to be much less expensive and
the number of new topologies per step depends strongly on the amplitude of the
displacement during the previous KMC step. Small displacements observed during
flickers usually result in less than 10 new topologies while diffusive events
can generate up to 140 new topologies in a single step.

The distribution of activation barriers is similar to the one found in previous
ART studies \cite{Valiquette:2003:125209,Kallel:2010:045503}. Although it is
continuous over a wide range of activation energies, the kinetics is dominated
by non-diffusive low energy barriers. Contrary to crystalline systems, where a clear
energy threshold separates diffusive from non-diffusive events, it is necessary
to fix the basin-threshold somewhat arbitrarily in the mean-rate method. Here we
chose a cutoff of 0.3 eV for a simulation temperature of 300 K. Therefore,
events associated with timescale of 16~ns or less are averaged over and the
internal dynamics of these events is ignored. This is acceptable, as we are
interested in simulations on the time scale of microseconds or more.

Figure~\ref{asi_plot} (bottom) shows the evolution of the total configurational
energy as a function simulated time for a simulation of 3360 k-ART
steps. Since we started from a very well-relaxed configuration, the
proportion of flickers is important but the system still manages to
relax by more than 6 eV over a 12 $\mu$s simulation. While the initial
relaxation with the modified Stillinger-Weber potential leaves the
system perfectly coordinated, an average of 0.8 at.\% defects are
created in a few KMC steps. This concentration is relatively constant
throughout the simulation. Moreover, almost all the low barrier events
involve coordination defects.  Defect migration events are hard to
characterize since local atomic motions can affect the existence of
low energy defects up to the third nearest neighbor distance.  This can cause
some defects to disappear while creating new ones.

The 4 eV drop at 10 $\mu$s is initiated by a bond switching event
of two four-fold coordinated neighboring atoms with a barrier of 0.28
eV. This allows for one atom to get rid of a highly strained bond,
resulting in a energy drop of 0.84 eV. This event is then followed by
a succession of 84 smaller relaxation events involving mostly 
spontaneous creation or destruction of low-energy coordination
defects. The configuration eventually ends up in a lower energy basin
where flickers again dominate. The average defect population goes from
0.8 to 1.4 at.\% during the entire process.

K-ART in a-Si can be compared, on short times, with molecular
dynamics. Using the MD software {\sc LAMMPS} \cite{Plimpton:1995:1,*LAMMPS} with our
a-Si model, we launch a 10-ns simulation at 300 K with a timestep of 2
fs. At regular intervals, a configuration is frozen and relaxed into
the nearest local minimum using steepest descent in order to compare
with our k-ART simulation. Results (not shown) confirm that almost no
deformation takes place on this short time scale and both MD and k-ART
display atomic displacement of the same amplitude.

Total simulation time for this system is significant and Fig.~\ref{asi_plot}
(top) shows the evolution of simulation time as a function of computer time for
a code running on a single 2.66 GHz Intel Xeon X5550 CPU starting from a preconstructed
catalog. The use of a parallel version of the algorithm coupled with a more
extensive catalog is expected to reduce considerably the computational efforts
for this simulation. Already, however, we see that k-ART can be a useful tool for these complex systems.

\section{Conclusion}

In this paper, we present in details the kinetic ART algorithm, a versatile
self-learning on-the-fly off-lattice kinetic Monte Carlo method. This method
couples ART nouveau~\cite{Malek:2000:7723}, a very efficient non-biased
open-ended algorithm for finding transition
states~\cite{Marinica:2011:094119,Machado-Charry:2011:034102}, with a topological
classification of events based on {\sc nauty}, a powerful packaged developed by
McKay~\cite{McKay:1981:45}.

Kinetic ART constructs a reusable event catalog that improves the efficiency of
the algorithm over time. Events are stored as generic events coupled to a given
topology. To fully include elastic deformations, the lowest-energy barriers are
separately relaxed to specific events to fully account for geometrical and
elastic deformations. By construction, the algorithm also automatically
identifies cases when the topology does not correspond to a single geometry,
ensuring that the basic approximations are valid for all events. For efficiency,
k-ART also includes local force calculations, allowing sub-linear scaling with
system size, and an exact handling of flickers extended from the mean-rate
method \cite{Puchala:2010:134104}. Other acceleration techniques, such as
parallel handling of event relaxation and generation, are also implemented in
the current version of the k-ART package.

To demonstrate k-ART's versatility, we applied the algorithm to three problems:
vacancy-interstitial annihilation in c-Si, interstitial diffusion in
Fe, and relaxation of a-Si. Clearly, the algorithm, although slower than
standard KMC, can handle accurately complex systems with many tens of thousands
of topologies much faster than MD, opening the possibility of studying problems
that have long remained out of reach of simulation.

\appendix*
\section{The Mean Rate Method}
\label{sec:mrm}
Following Puchala \etal\ \cite{Puchala:2010:134104}, the system is
separated in transient states and absorbing
states. To determine the probability to exit the basin to state $x$,
we calculate the transistion probability matrix $\te T$, with
components
\begin{equation}
  \label{eq:transmat}
  T_{ji}=\frac{R_{i\rightarrow j}}{\sum_k R_{i\rightarrow
      k}}=\tau^1_iR_{i\rightarrow j}, 
\end{equation}
where $R_{i\rightarrow j}$ is the rate going from basin state $i$ to
basin state $j$, and the summation is over all basin and exit states
$k$.  $\tau^1_i$, the reciprocal of the sum of all rates leaving state
$i$, is the mean residence time in state $i$ each time it is
visited. The occupation probability vector of all basin states after
in-basin jump $m$ (and before $m+1$), $\ve{\Theta}(m)$ is thus given
by repeated application of $\te T$ to the initial occupation
probability $\Theta_i(0)=\delta_{is}$, where $s$ is the starting
state. The sum of the occupation probabilities over all possible
number of jumps gives the average number each basin state is visited:
\begin{equation}
  \label{eq:occprobsum}
  \ve{\Theta}^\text{sum}=\sum_{m=0}^{\infty} \te{T}^m \ve{\Theta}(0) =
  (\mathbbm{1} -  \te{T})^{-1}\ve{\Theta}(0),
\end{equation}
from which the mean residence time in basin state $i$ before leaving
the basin can be calculated:
\begin{equation}
  \label{eq:meanrestime}
  \tau_i=\tau^1_i \Theta_i^\text{sum}.
\end{equation}
These residence times are then used to accelerate the basin exit rates
from basin state $i$ to exit state $j$ according to
\begin{equation}
  \label{eq:accrate}
  \langle R_{i\rightarrow j}\rangle= \frac {\tau_i}{\sum_k \tau_k} R_{i\rightarrow j},
\end{equation}
with $k$ summing over all basin states. The next KMC step is then
determined using standard KMC rules, using these accelerated rates.

In contrast to the first passage time analysis (FPTA)
\cite{Novotny:1995:1,Puchala:2010:134104}, the mean rate method is
computationally much simpler, as it requires a single matrix
inversion to calculate the modified rates, after which the ordinary
KMC rules apply. This comes at a cost: There is no correlation between
the randomly determined residence time and the selected exit
state. Puchala \etal\ find \cite{Puchala:2010:134104} that in measuring
average quantities after many steps, both MRM and FPTA yield the same
results.

\begin{acknowledgments}

  This work has been supported by the Canada Research Chairs program
  and by grants from the Natural Sciences and Engineering Research
  Council of Canada (NSERC) and the \textit{Fonds Qu\'eb\'ecois de la
    Recherche sur la Nature et les Technologies} (FQRNT). We are
  grateful to \textit{Calcul Québec} (CQ) for generous allocations of computer
  resources. %

\end{acknowledgments}

\bibliography{makros,papers,kART_2011}

\end{document}